# A General Way to Control the Reversal of Near Field Optical Binding Force between Plasmonic or Dielectric Dimers

Hamim Mahmud Rivy[1+], M.R.C. Mahdy[1+*], Ziaur Rahman Jony[1], Nabila Masud[1], Sakin S Satter[2], and Rafsan Jani[1]

[1] *Department of Electrical & Computer Engineering, North South University, Bashundhara, Dhaka, 1229, Bangladesh.*

[2] *Department of Electrical & Electronic Engineering, Dhaka University, Dhaka, Bangladesh.*

+ Equal contribution

*Corresponding Author: mahdy.chowdhury@northsouth.edu

**Controlling the near field optical binding force can be a key factor for particle clustering, aggregation and localized surface plasmon sensors. So far there is no generic way to reverse the near field optical binding force for plasmonic or dielectric nano-dimers of distinct shapes (cube, cylinder, ring, sphere). In this article, for both plasmonic and dielectric objects, we have demonstrated a general way to control the reversal of near field binding force for different shaped dimer sets. The force reversal is achieved by simple breaking of symmetry, considering the nanoparticles are half or less than half immersed in an inhomogeneous dielectric background, i.e. at air-water interface. Such reversals have been explained based on Fano resonance, interference fields, unusual behavior of optical Lorentz force and image charge theory. However, if the dimer set is placed over a dielectric interface or fully inside a homogeneous medium, the sign of binding force does not reverse. Our proposed configuration provides a generic mechanism of archiving binding force reversal for both plasmonic and dielectric objects, which can be verified by very simple experimental set-up.**

**Introduction**

Optical manipulation based on optical gradient, scattering and curl force has been widely applied and investigated in the fields of physics [1-7], physical chemistry and plasmonics [8, 9], biological science [10,11], and nano science [12]. To manipulate the micro objects cooling was later introduced [13], though cooling is not an option for studying dynamics that occur at room temperature or in liquid medium [14]. On the other hand, by using holographic or scanning optical tweezers [15, 16], multiple particles have been successfully manipulated. However, such techniques require widely distributed laser intensity and the structured light fields [16]. An alternative way can be the increase of laser intensity for better controlling of micro particles [17]. However, the increase of light intensity may cause damage to the sample, such as biological objects [18].

There is a different kind of optical force other than gradient, scattering and curl force; known as the optical binding force [19-27], which can also be used as a powerful tool to precisely control the mutual attraction and repulsion between different objects for colloidal self-assembly, crystallization, and so on [27] without the need for structured light fields [16] or high intensity light fields [17]. Most of the works on optical binding force have been investigated considering the inter-particle distance as several micrometers [28-32]. In contrast, an emergent and powerful topic in the field of optical binding, that is still very new and steadily growing, is near field optical binding force between plasmonic dimers [8, 19, 23, 33-37], which can be very useful for particle clustering and aggregation [37].

The science dealing with the collective oscillations of the conduction electrons of plasmas is known as plasmonics, which is a powerful tool to incorporate optics into nanoscience and nanotechnology. The promising applications of plasmonic hybridization and Fano resonances have been investigated in improved sensitivity of the resonance, bio sensing, surface-enhanced Raman scattering, plasmon-induced transparency and many others [38-44]. In contrast, just few works have investigated the behavior of near field optical binding force [8, 19, 23, 33-37] and almost all of them have focused the near field binding force of plasmonic objects instead of dielectric objects. Among those reports, even fewer works have reported another novel approach of near field optical binding force: achieving the reversal of near field binding force for *plasmonic dimers* [8, 33-37]. The adjustment of attraction and repulsion between very closely placed nano-dimers is known/ termed as the reversal of near field optical binding force. This novel area is even younger than some of the recent popular areas of optical manipulation: the reversal of optical scattering force or passive tractor beams [45-48], active tractor beams [49, 50] and optical lateral force [9, 51, 52].

So far, reversal of near field optical binding force has been reported for plasmonic heterodimers in very few works [8, 33-37]: between plasmonic nanorod heterodimers [33], between disc-ring hetero structures

[34], between plasmonic spherical heterodimers [8, 36, 37].Only for cube shaped plasmonic *homodimers*, controlled attraction and repulsion of such near field binding force has been reported when they are placed over plasmonic substrate [35]. However, there is no generic way to reverse the near field optical binding force for both plasmonic and dielectric nano-dimers (either homodimers or heterodimers) of distinct shapes (cube, cylinder, ring, sphere). Notably, the wide area of near field optical binding force for dielectric nano-dimers, which can open a novel way to manipulate biological objects, is still remain unexplored.

In this article, for both plasmonic and dielectric objects, we have demonstrated a general way to control the reversal of near field binding force for different shaped (cube, cylinder, ring, sphere) dimer sets (either homodimers or heterodimers). The force reversal is achieved by simple breaking of symmetry, considering the nanoparticles are half or less than half immersed in an inhomogeneous dielectric background, i.e. at air-water interface. We have shown the clear connection of plasmonic binding force reversal with induced Fano resonance. However, according to our investigation, reversals of optical binding force occurs not only for plasmonic nano-dimers but also for dielectric nano-dimers where Fano resonance does not occur. As a result, one of our notable conclusions is that: though Fano resonance can be one of the key parameters for the reversal of near field optical binding force [33-35], it is not a generic/common feature for the reversal of near field optical binding force. In this work, such reversals have also been explained based on the unusual behavior of optical Lorentz force, interference fields and image charge theory. We have also proposed a simple way to control the magnitude and the shifting of such binding force reversal wavelength by using only the permittivity of lower background medium as the controlling parameter. It should be noted that if the dimer set is placed over a dielectric interface or fully inside a homogeneous medium, the sign of binding force does not reverse.

Previous works [8, 33-37] have demonstrated the strong dependency of near field optical binding force reversal on different parameters of plasmonic objects: particle size, shape, inter particle distance etc. But our proposed current configuration provides not only much relaxation over those parameters but also a generic mechanism of archiving binding force reversal for both plasmonic and dielectric objects, which can be verified by very simple experimental set-up.

**Methods**

Fig.1 illustrates a schematic diagram of our proposed simple set up. Where Fig.1(a) shows cube shaped nanoparticle homodimer half immersed in an inhomogeneous dielectric background under plane wave illumination. The nanoparticles are placed d=100 nm apart from each other (surface to surface). Fig.1(d), (g) and (j) show the similar set up for cylinder, ring and spherical shaped nanoparticle homodimers half immersed in the underneath background medium respectively. In this article we have not only studied the behavior of nanoparticles half immersed but also for two-third immersed and one-third immersed in the inhomogeneous background medium [supplement S1 shows the nanoparticle dimers immersed one-third S1: Fig.1(a)-(d) and two-third S1: Fig.1(i)-(l) in the underneath background medium]. The dimensions of the nanoparticle homodimers are (i) Cube (length 150 nm each), (ii) Cylinder (radius 150 nm and height 150 nm), (iii) Ring (inner radius 30 nm, outer radius 75 nm and height 150 nm) and (iv) Sphere (radius 75 nm). The nanoparticle on the left is considered as nanoparticle-1 and the right one is nanoparticle-2. For most of the cases, we have used 1.33 (water) as the background medium Refractive index(RI). The real and imaginary part of the permittivity of silver is taken from standard Palik (0-2µm) data [53,54] (these criteria are in the best agreement with the FDTD fitting model in [53] for full wave simulation).We have varied the wavelength from 300nm to 1200 nm and considered simple *x*-polarized plane wave $E_x = e^{jkz}$ propagating towards $-z$ direction.

Throughout the paper "all four cases" represents the cubic, cylindrical, ring shaped and spherical homodimers. The forces calculated outside the volume of the nanoparticles defined as 'exterior' or 'outside' forces and those which are calculated inside are referred to as 'interior' or 'inside' forces.

The 'outside optical force' [55] is calculated by the integration of time averaged Minkowski stress tensor [9,55,56] at r=$a^+$ employing the background fields of the scatterer of radius *a*:

$$\left\langle \boldsymbol{F}_{\text{Total}}^{\text{Out}} \right\rangle = \sum_{(j)} \oint \left\langle \bar{\bar{\boldsymbol{T}}}_{(j)}^{\text{out}} \right\rangle \cdot d\boldsymbol{s}_{(j)}$$
$$\left\langle \bar{\bar{\boldsymbol{T}}}_{(j)}^{\text{out}} \right\rangle = \frac{1}{2} \text{Re}[\boldsymbol{D}_{\text{out}(j)} \boldsymbol{E}_{\text{out}(j)}^* + \boldsymbol{B}_{\text{out}(j)} \boldsymbol{H}_{\text{out}(j)}^* - \frac{1}{2} \bar{\bar{\boldsymbol{I}}} (\boldsymbol{E}_{\text{out}(j)}^* \cdot \boldsymbol{D}_{\text{out}(j)} + \boldsymbol{H}_{\text{out}(j)}^* \cdot \boldsymbol{B}_{\text{out}(j)})]$$ (1)

Where *j*=1 (upper background) or *j*=2 (lower background) represents the exact background region (either upper medium or lower medium) sharing interface with the object, [cf. 1st column in our Fig. 1]. $\boldsymbol{E}$, $\boldsymbol{D}$, $\boldsymbol{H}$ and $\boldsymbol{B}$ are the electric field, displacement vector, magnetic field and induction vector respectively, $\langle \ \rangle$ is the time average and $\bar{\bar{\boldsymbol{I}}}$ is the unity tensor and 'out' represents the exterior total field of the scatterer.

On the other hand, based on the Lorentz force, the total force (surface force and the bulk force [6, 35, 36, 57]) can be written as:

$$\left\langle \boldsymbol{F}_{\text{Total}} \right\rangle = \left\langle \boldsymbol{F}_{\text{Volume}} \right\rangle = \sum_{(j)} [\left\langle \boldsymbol{F}_{\text{Bulk}(j)} \right\rangle + \left\langle \boldsymbol{F}_{\text{Surf}(j)} \right\rangle] = \sum_{(j)} [\int \left\langle \boldsymbol{f}_{\text{Bulk}(j)} \right\rangle dv_{(j)} + \int \left\langle \boldsymbol{f}_{\text{Surface}(j)} \right\rangle] ds_{(j)}$$ (2)

Where

$$\langle f_{\text{Surface}(j)}\rangle = \sigma_e E^*_{avg(j)} + \sigma_m H^*_{avg(j)}$$

$$= \{\epsilon_o(E_{\text{out}(j)} - E_{\text{in}(j)})\cdot \hat{n}\}\left(\frac{E_{\text{out}(j)} + E_{\text{in}(j)}}{2}\right)^* + \{\mu_0(H_{\text{out}(j)} - H_{\text{in}(j)})\cdot \hat{n}\}\left(\frac{H_{\text{out}(j)} + H_{\text{in}(j)}}{2}\right)^*, \quad (3)$$

$$\langle f_{\text{Bulk}(j)}\rangle = \frac{1}{2}\text{Re}[\varepsilon_0(\nabla\cdot E_{\text{in}(j)})E^*_{\text{in}(j)} + \mu_0(\nabla\cdot H_{(j)})H^*_{\text{in}(j)}] - \frac{1}{2}\text{Re}[i\omega(\varepsilon_{s(j)} - \varepsilon_b)\{E_{\text{in}(j)}\times B^*_{\text{in}(j)}\}$$
$$+ i\omega(\mu_{s(j)} - \mu_b)\{D^*_{\text{in}(j)}\times H_{\text{in}(j)}\}] \quad (4)$$

$f_{\text{Surface}}$ and $f_{\text{Bulk}}$ are the surface force density and bulk force density respectively. $f_{\text{Surface}}$ is calculated just at the boundary of a scatterer and $f_{\text{Bulk}}$ is calculated from the interior [6,35,36,57]. $\varepsilon_s$, $\mu_s$ are the permittivity and permeability of the nanoparticle respectively. $\varepsilon_b$ and $\mu_b$ is for the background. $\hat{n}$ is an outward pointing normal to the surface. 'in' represents the interior fields; 'avg' represents the average of the field. $\sigma_e$ and $\sigma_m$ are the bound electric and magnetic surface charge densities of the scatterer respectively. As far as our concern, the Lorentz force dynamics of plasmonic particles has been only studied for plasmonic homodimers over plasmonic substrate [35] and on axis and of axis heterodimers [36]. But it has not yet been studied for plasmonic and dielectric homodimers half immersed in a mixture of inhomogeneous dielectric medium.

The 'external dipolar force' [1,3,5,8] (which has also been described as Lorentz force in [8]) is quite different than the Lorentz force [6,35,36,57] defined in our Eqs (2) - (4). Though sometimes quasi static analysis (i.e. dipolar force ) leads to wrong conclusion (for example- some cases are discussed in refs [8], [18] ); the agreement of Lorentz volume force [6,35,36,57] and external ST method [4,7,8,9,55,56] based on full electrodynamic analysis [ which is considered for all the force calculations in this article] should lead to the consistent result for realistic experiments. All the numerical calculations are done in full wave simulations [53] with 3D structures.

The difference of the scattering part [cf. Eq (4)] or bulk part of the total Lorentz force on a plasmonic object should describe the relative bulk force experienced by the optical molecule:

$$\text{Del F}_{\text{Bulk}} = \int[\langle f_{\text{Bulk }(B)}\rangle dv_{(B)}] - \int[\langle f_{\text{Bulk }(S)}\rangle dv_{(S)}] \quad (5)$$

Here; subscript (B) and (S) represent: bigger object and smaller object respectively. At the same time the difference of the gradient part [which originates from induced surface charges; cf. Eq (3)] of the total

Lorentz force on a plasmonic object should describe the relative surface force experienced by the optical molecule:

$$\text{Del } F_{Surf(x)} = \int [\langle f_{Surface(B)} \rangle ds_{(B)}] - \int [\langle f_{Surface(S)} \rangle ds_{(S)}] \quad (6)$$

It should be noted that: $F_{Bind} = (F_B - F_S) = \text{Del } F_{Bulk} + \text{Del } F_{Surf}.$ (7)

**Results and discussions**

In supplement S2, we have shown that if the nanoparticle homodimers are kept in vacuum no reversal of near field optical binding force occurs. Even if they are kept above a dielectric substrate no binding force reversal is observed. To overcome this problem, we have employed a set up where nanoparticle homodimers are half immersed or less than half immersed in an inhomogeneous dielectric background. As shown in Fig.1 where the nanoparticles are half immersed in the lower background reversal of near field optical binding force occurs near Fano dip.

The extinction spectra of Fig 1. shows the bright resonant mode, dark resonant mode and Fano dip for distinct shaped (cube, cylinder, ring and sphere) nanoparticles half immersed in an inhomogeneous dielectric background. The table below shows the wavelength at which bright resonance mode, Fano dip and dark resonance mode occurs for cube, cylinder, ring and sphere shape nanoparticle homodimers.

Table 1

| Shape | Bright(nm) | Fano dip(nm) | Dark(nm) |
|---|---|---|---|
| Cube | 427 nm | 467 nm | 721 nm |
| Cylinder | 407 nm | 446 nm | 630 nm |
| Ring | 427 nm | 450 nm | 665 nm |
| Sphere | 327 nm | 374 nm | 611 nm |

The binding force is represented by $F_{bind(x)} = (F_{1x} - F_{2x})$. Positive $F_{bind(x)}$ means attractive and negative $F_{bind(x)}$ means repulsive forces respectively. In all four cases, $F_{bind(x)}$ reaches maximum attractive force near bright resonant mode as shown in the binding force curve in Fig.1. Local attractive maximum binding force occurs between Fano dip and dark resonant mode, whereas local repulsive maximum occurs between Fano dip and bright resonant mode. The extinction spectra and binding force curve of Fig.1 reveals that the binding force reverses near dark resonant mode, Fano dip and between Fano dip and bright resonant mode.

The detailed mechanism for such force reversal and the effect of dielectric medium is discussed in the next part.

The Fano resonance in plasmonic nanoparticle originates from the destructive interference of the bright and dark plasmon resonant mode. It was previously shown [35] that when two Ag homodimers are placed over a silicon or glass substrate or other dialectic [ as illustrated in supplement S2 for water substrate] substrate, Fano dip (or weak Fano dip) is observed but no reversal of optical binding force occurs. But when nanoparticles are immersed half in the background, near field optical binding force reversal occurs. This phenomenon can be clearly understood by image charge model [37, 58]. When the nanoparticle dimers are placed over a dielectric substrate the image charge induced interaction with the substrate is weaker. As a result, no reversal of near field optical binding force occurs. When the nanoparticle dimers are half immersed in the underneath background, the interaction with the image charge leads to stronger interaction between bright and dark resonant mode, which leads to stronger Fano resonance. The strong Fano resonance causes the binding force reversal near Fano dip.

As observed from Fig. 1(c), (f), (i) and (l), force reversal occurs in different wavelength regions other than Fano dip. In addition, when the nanoparticles are placed above the dielectric (lower) background, Fano dip is observed, but no binding force reversal is found [supplement S2]. So, it clearly indicates that Fano resonance cannot be the only reason for binding force reversal.

From Fig. 2 (a)-(d), it is observed that near dark resonant mode, the Del $F_{surf(x)}$ and Del $F_{bulk(x)}$ reach their local maximum negative and positive value respectively. Binding force reversal occurs near dark resonant mode where $F_{bind(x)}$ becomes negative to positive (repulsive to attractive) due to slight dominance of Del $F_{bulk(x)}$. Another force reversal occurs near Fano dip where an interesting phenomenon is observed. Del $F_{surf(x)}$ and Del $F_{bulk(x)}$ both forces reverse their sign near Fano dip. It is observed in all four cases (cube, cylinder, ring and sphere) that Del $F_{bulk(x)}$ becomes positive to negative and Del $F_{surf(x)}$ becomes negative to positive near Fano dip. Near Fano dip region, the near field optical binding force $F_{bind(x)}$ reverses again due to the dominance of Del $F_{bulk(x)}$. $F_{bind(x)}$ becomes repulsive from attractive force. It indicates that bulk force plays very effective role in binding force reversal in plasmonic homodimers. Another force reversal occurs at the wavelength where Del $F_{bulk(x)}$ reaches its negative maximum value and the force reversal occurs due to the dominance of Del $F_{surf(x)}$. Near bright resonant mode: the resultant attractive binding force reaches the maximum and it is dominated by the Del $F_{surf(x)}$, where the bulk force is almost zero. The maximum attractive binding force originates near bright resonant mode is due to the strong interaction between the image charge induced in the background index and the surface plasmon of the nanoparticles.

It is usually believed that plasmonic forces mostly arise from the surface force/polarization induced charges [59, 60]. But our study based on Lorentz force suggests a notably different proposal for half immersed plasmonic dimers (as bulk part of Lorentz force plays a vital role for the reversal of near field optical binding force).

**The effect of permittivity and different immersed conditions**

In Fig. 3(a)-(h) the extinction spectra and corresponding force curve is shown for different permittivity of the underneath background dielectric medium. We have varied the Refractive Index of the background medium from (1.33-3). With increasing permittivity, the resonance peak of both bright resonant mode and dark resonant mode red-shifts. The bright resonant mode of extinction spectra red shifts more with a little increase in permittivity but the dark resonant mode shows a very small red shift. In all four cases, near bright resonance mode the electric field distribution is localized at the air-water dielectric interface but for dark resonant mode Electric field is concentrated at the top of the nanoparticles (air side) [electric field distribution near bright, Fano dip and dark resonant mode is given in supplement S3 for all four cases]. With increasing the permittivity, dielectric screening factor ($\varepsilon_s$ -1)/ ($\varepsilon_s$ + 1) increases, which leads to a much stronger interaction between delocalized surface plasmon and the induced image charge during bright resonant mode than dark resonant mode ($\varepsilon_s$ is considered as the dielectric constant of the background dielectric medium). As a result, bright resonant mode red shifts more than the dark resonant mode and the splitting between the two-resonant mode increases with increasing the permittivity of the background medium.

The extinction spectra of Fig.3 (a)-(d) show that increasing permittivity leads to an enhanced Fano resonance and greater red shift in Fano resonance in the extinction spectra. Large image charge and strong interaction (between the image charge and delocalized surface plasmon) also originates from the background with high permittivity. This strong interaction between large image charge with bright and dark resonant mode creates enhanced Fano resonance. The width of the Fano resonance also increases due to splitting between the bright and dark resonance mode. As bright resonant mode red shifts more than the dark resonant mode, the splitting between bright and dark resonance mode increases. The increasing admixture of bright and dark resonance mode also affects the width of the Fano resonance. As a consequence, Fig.3 (e)-(h) show repulsive binding force is observed over large spectral region with greater magnitude for high permittivity lower background medium shown.

As observed in Fig.3 (e)-(h), with increasing the permittivity of the underneath background medium the magnitude of the attractive binding force increases near dark resonance mode.

And also, high permittivity background leads to a strong mixture of bright and dark resonant mode. As discussed in ref. [58], the image charge induced by electromagnetic field of a real dipolar nanoparticle plasmon will have a quadrupolar component across the nanoparticle; and image charge induced by the quadrupolar component will have a dipolar component across the nanoparticle. It means that, for high permittivity dielectric background the dark resonant mode is actually a mixture of image charge induced bright resonant mode component and the dark mode component. Similarly, bright resonant mode is actually a mixture of image charge induced dark resonant mode component and the bright resonant mode component. As a result, a net dipole moment arises near the dark resonant mode. Basically due to that induced dipole moment near the dark resonant mode, the magnitude of the attractive binding force increases as illustrated in Fig.3(e)-(h).

In Fig.4 we have shown the extinction spectra for three different immersed conditions for plasmonic nanoparticles and the behavior of near field binding force with respect to the conditions. Particle is immersed 50nm, 75nm and 100nm under the background medium are called one-third immersed (50nm), half immersed (75nm) and two third immersed (100nm) respectively [Fig.1 in supplement S1 shows (a)-(d) for one-third, (e)-(h) show half and (i)-(l) for two third immersed condition]. The extinction spectra of Fig.4 (a)-(d) reveal that when one-third of the homodimers (all four cases) are immersed in the water the dark resonant mode is more visible and have more distinct peaks. On the other hand, when the homodimers are immersed in two-third under water, it is very hard to detect the dark resonant mode. Fig.4 (e)- (h) reveal that when the particles are one-third and two-third immersed in the underneath background medium no reversal of near field optical binding force occurs near Fano dip.

Such phenomenon can be explained by electric field distribution for different immersed condition [The electric field distribution of different immersed condition during bright, dark and Fano resonance is given in supplement S4]. As discussed earlier, electric field distribution near the dark mode is oriented towards vacuum and near bright mode towards the air water. When the particles are half immersed in water, a strong mixture of bright and dark resonant mode leads to stronger Fano resonance. On the other hand, when the particle is immersed one-third under the background medium Fano resonance is weaker due to the distance between the interface and the top of the particle increases which leads to weaker Fano resonance. Due to weaker Fano resonance no force reversal is observed near Fano dip. But force reversal can be achieved with increasing the permittivity of the background index [more details on supplement S5]. When the particle is two-third immersed in water, distance between the interface and the top decreases and the electric field is oriented towards the interface. As a result, no distinct dark mode is visible during two-third immersed condition.

So, the force reversal can be achieved by simple breaking of symmetry, considering the nanoparticles are half (or less than half) immersed in an inhomogeneous dielectric background, i.e. at air-water (or other air-liquid interface where the liquid has higher refractive index than water).

**Force reversal for dielectric nanoparticle dimer**

The wide area of near field optical binding force for dielectric nano-dimers, which can open a novel way to manipulate biological objects, is still remained unexplored. In this section, we have demonstrated that such reversals are also possible for dielectric nano-dimers in very broad wavelength regions, though the magnitude of force is hundred times smaller than the force magnitude of plasmonic dimers.

In our analysis, we have used RI=1.45 for cubic, cylindrical, ring and spherical dielectric homodimers with same dimensions used in plasmonic nanoparticles discussed above. From Fig.5 (e)-(h) it is observed that reversal of near field optical binding force occurs for cubic, cylindrical and spherical shaped homodimers, except for ring shaped nanoparticle homodimer. The extinction spectra of Fig.5 reveals that the extinction coefficient decreases with increasing wavelength. But it was completely different for plasmonics, where different resonance picks are observed for plasmonic homodimers. As far as we know, it is quite unusual finding of near field optical binding force in dielectric homodimers which has not been studied in any literature yet. As illustrated in Fig.5, it is observed that Del $F_{surf(x)}$ increases with increasing the wavelength where Del $F_{bulk(x)}$ decreases. The binding force reversal occurs due to positive Del $F_{bulk(x)}$ and rapidly decreasing Del $F_{surf(x)}$ and their maximum value occurs due to the effectiveness of both Del $F_{surf(x)}$ and Del $F_{bulk(x)}$. But it is noticed from Fig.5(g) no reversal occurs in the ring-shaped homodimers and always remains positive due to a strong positive bulk force and a weak negative surface force.

**CONCLUSION**

In conclusion, we have proposed and investigated a simple configuration to control the reversal of near field optical binding force for different shaped and sized plasmonic and dielectric nano-dimers immersed in an inhomogeneous dielectric background; i.e. at air-water interface. We have also verified the robustness of our proposed set-up by investigating the behavior of force reversal for different immersed conditions/ situations. The force magnitude and shifting wavelength of such force reversal over different wavelengths can be controlled/tuned by changing the permittivity of the background medium. Notably, the wide area of near field optical binding force for dielectric nano-dimers, which can open a novel way to manipulate biological objects, was remained unexplored. We have demonstrated that such reversals are also possible for dielectric nano-dimers in very broad wavelength regions, though the magnitude of force is hundred times smaller than the force magnitude of plasmonic dimers. As most of the real world experimental set-ups involve liquid material as the background for optical manipulation, our proposed simple technique may open novel routes not only for particle clustering or aggregation but also for colloidal self-assembly, crystallization, and organization of templates for biological and colloidal sciences.

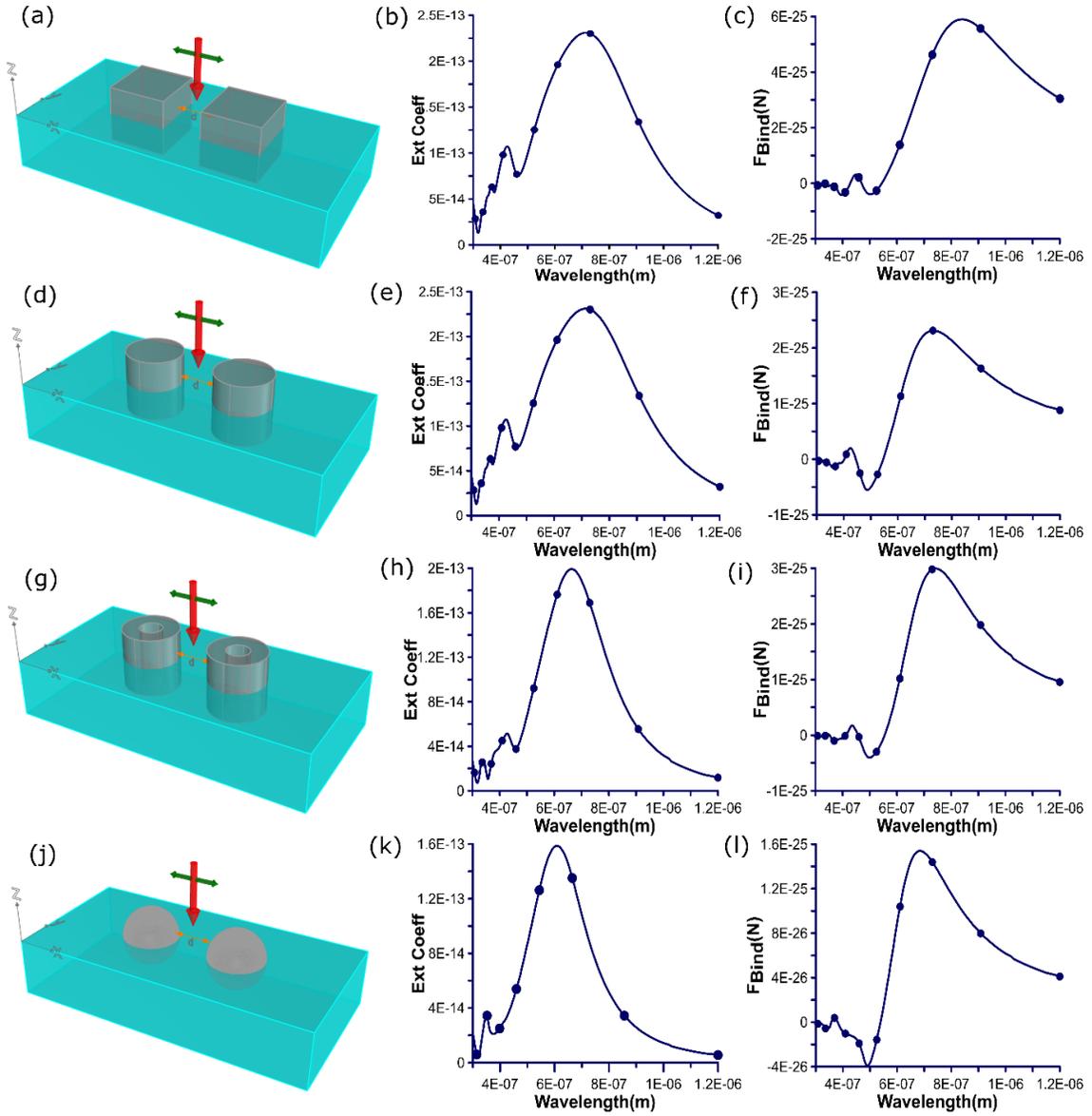

Fig. 1 (a) shows the cube nanoparticle homodimer placed in an inhomogeneous background (Refractive index=1.33), (b) & (c) show extinction spectra and $F_{bind(x)}$ for the nanoparticle cube homodimer respectively. Similarly, (d)-(f) represent the cylinder, (g)-(i) for ring and (j)-(l) for spherical shaped nanoparticle homodimer. Left side negative (x) denoted as particle 1 and right side denoted as particle 2. $F_{bind(x)}$ is $(F_{1x} - F_{2x})$ for each case. The wavelength at which bright resonant mode, Fano dip and dark resonant mode occurs is given in Table 1.

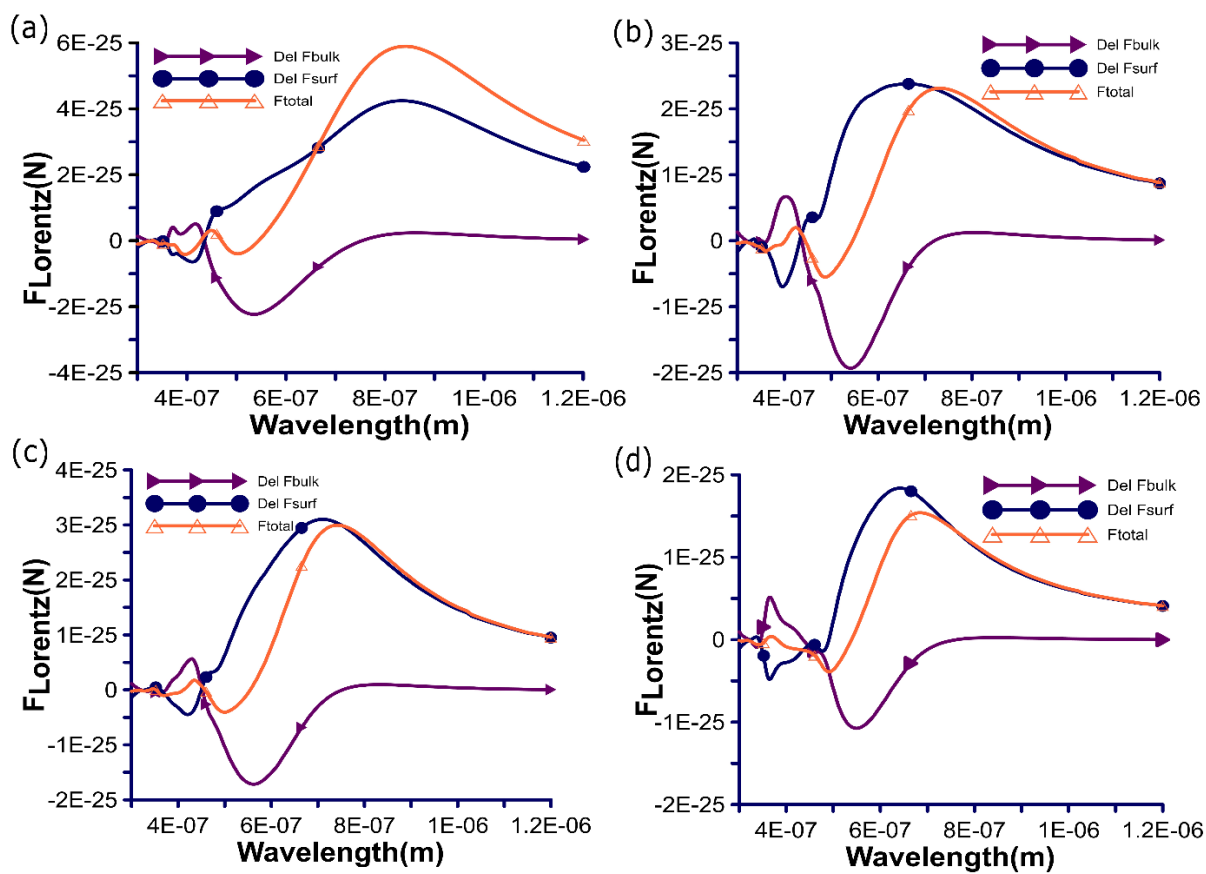

Fig. 2 Lorentz force components for the nanoparticle dimers immersed half in the background. Del $F_{surf(x)}$, Del $F_{bulk(x)}$ and binding $F_{bind(x)} = (F_{1x} - F_{2x})$ [cf. Eq. (5), (6) and (7) in main text] force components (a) cubic, (b) cylindrical, (c) ring shaped, (d) spherical plasmonic homodimers.

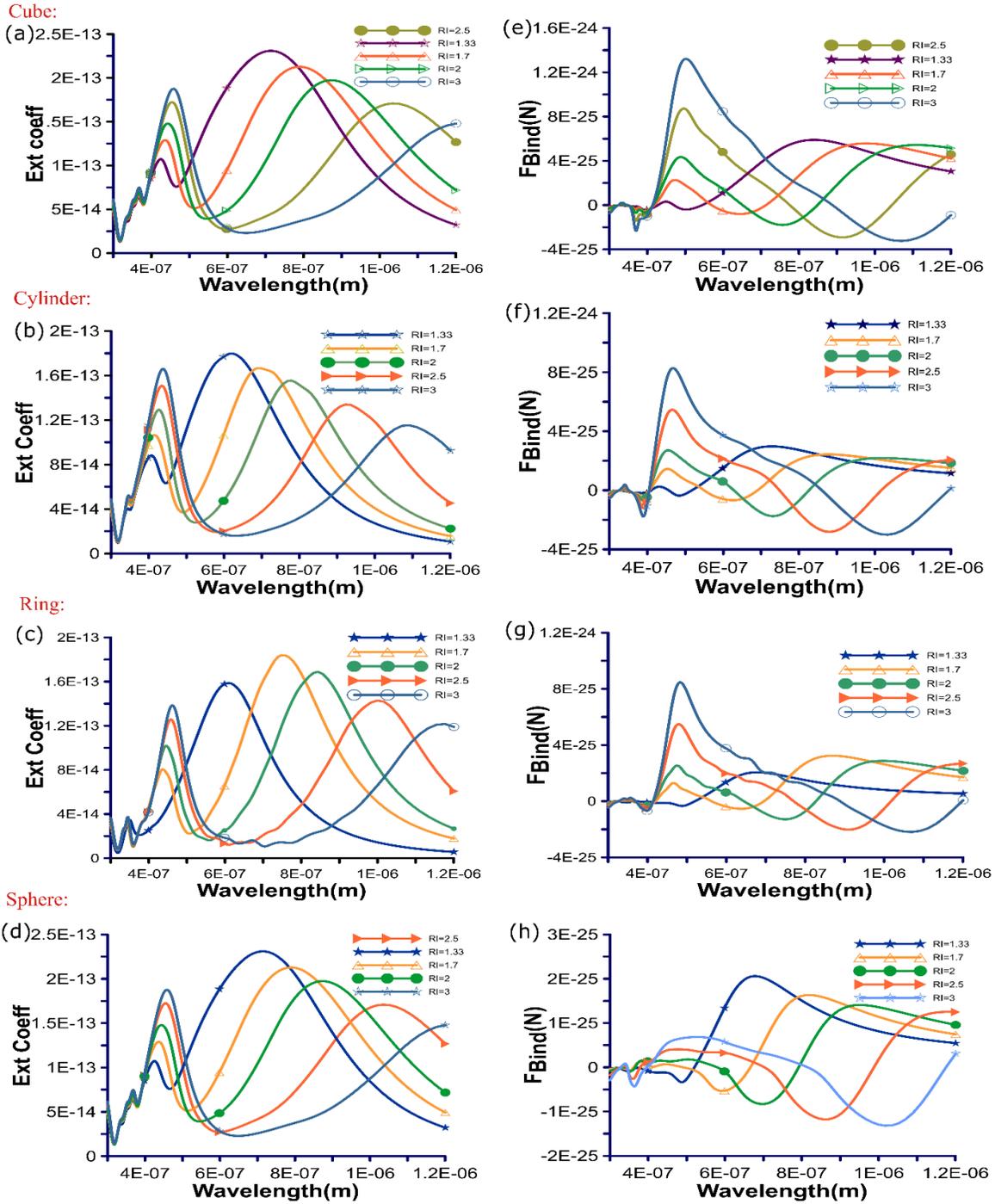

Fig. 3 **The** 1st colum represents the extinction spectra for different Refractive Index (RI=1.33,1.7,2,2.5,3 respectively) of the background medium and 2nd colum the corresponding force curve. (a) & (e) represent the extinction spectra and binding force curve for cube homodimers, (b) & (f) is for cylindrical homodimer, (c) & (g) for ring shaped and (d) & (h) represent the spherical homodimers respectively.

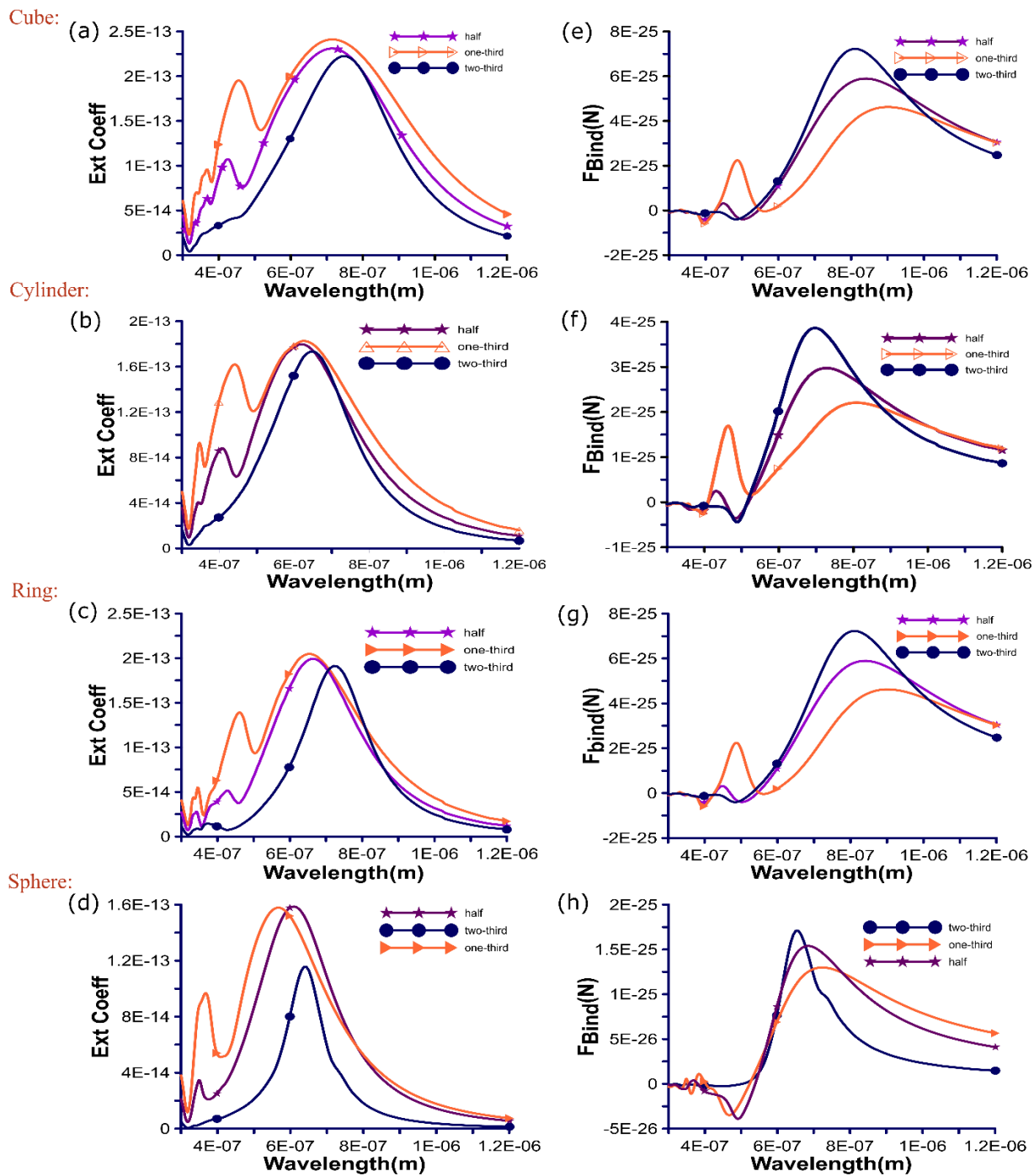

Fig. 4 Description of the extinction spectra (a)-(d) and corresponding force curve (e)-(h) are given for plasmonic homodimers for three different immersed condition. 1$^{st}$ row represents the extinction spectra and binding force curve cube, 2$^{nd}$ row is for cylinder, 3$^{rd}$ row is for ring and 4$^{th}$ one is for spherical shape respectively. A schematic diagram for one-third, half and two-third immersed condition is given in supplement S1.

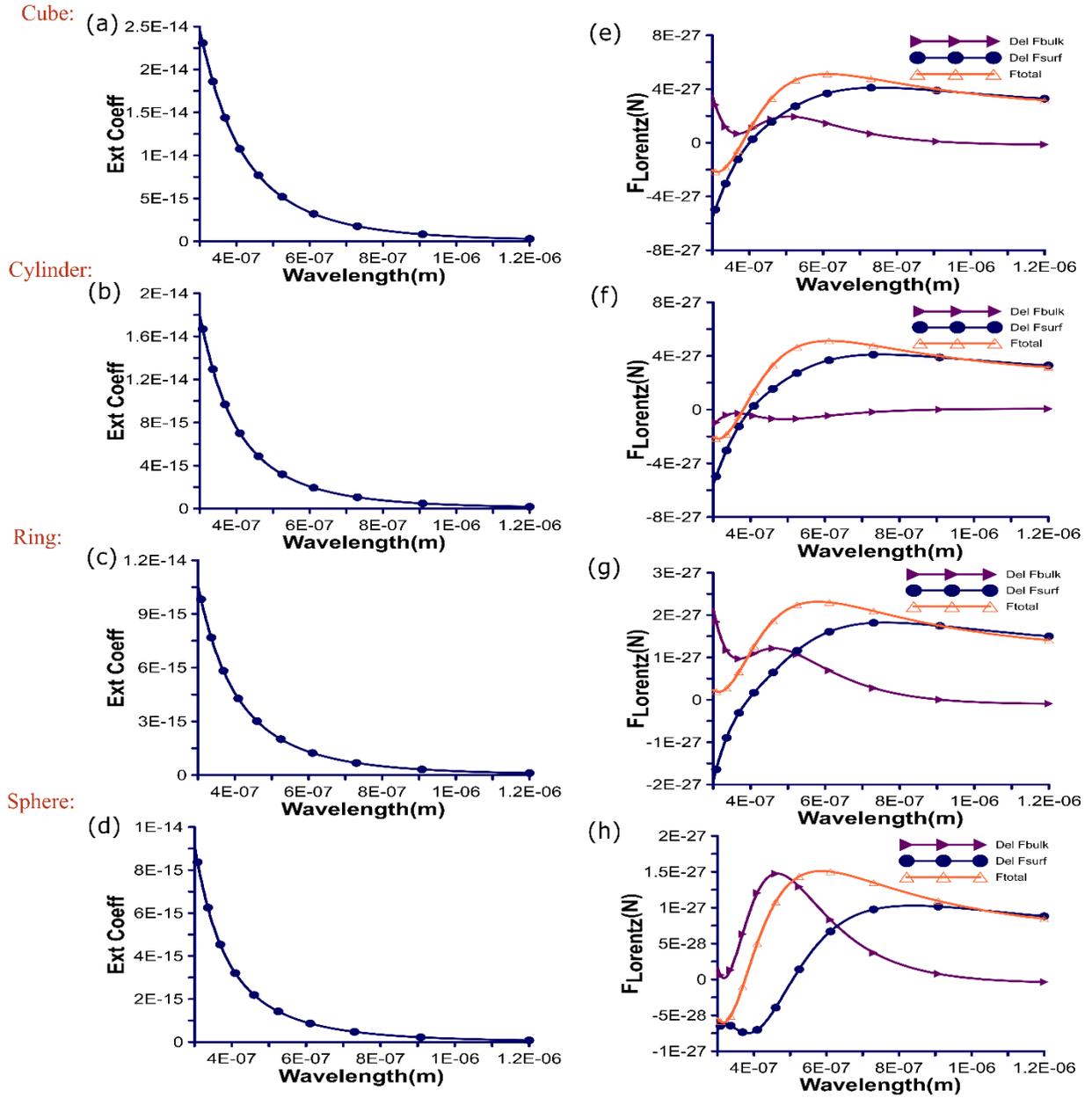

Fig. 5  Extinction spectra (a)-(d) and corresponding and Lorentz force components: (e)-(h) surface force, bulk force and binding $F_{bind(x)} = (F_{1x} - F_{2x})$ for dielectric homodimers. 1st row represents the extinction spectra and binding force curve cube, 2nd row is for cylinder, 3rd row is for ring and 4th one is for spherical shape respectively.